\documentclass[prl,amsmath,amssymb,twocolumn,superscriptaddress]{revtex4-2}

\usepackage{graphicx}
\usepackage{dcolumn}
\usepackage{bm}
\usepackage{color}

\begin{document}

\newcommand{\SIO}{SrIrO$_3$}
\newcommand{\onelayer}{Sr$_2$IrO$_4$}
\newcommand{\twolayer}{Sr$_3$Ir$_2$O$_7$}
\newcommand{\STO}{SrTiO$_3$}

\newcommand{\red}[1]{\color{red}{#1}}

\title{Coexistence of insulator-like paramagnon and metallic spin-orbit exciton modes in \SIO}

\author{E. Paris}
\email[]{eugenio.paris@psi.ch}
\affiliation{PSI Center for Photon Sciences, CH-5232 Villigen-PSI, Switzerland}

\author{W. Zhang}
\affiliation{PSI Center for Photon Sciences, CH-5232 Villigen-PSI, Switzerland}

\author{Y. Tseng}
\affiliation{PSI Center for Photon Sciences, CH-5232 Villigen-PSI, Switzerland}
\affiliation{Institute of Physics, National Yang Ming Chiao Tung University, Hsinchu 300093, Taiwan}

\author{A. Efimenko}
\affiliation{European Synchrotron Radiation Facility, 71 Avenue des Martyrs, 
38043 Grenoble, France}

\author{C. Sahle}
\affiliation{European Synchrotron Radiation Facility, 71 Avenue des Martyrs, 
38043 Grenoble, France}

\author{V. N. Strocov}
\affiliation{PSI Center for Photon Sciences, CH-5232 Villigen-PSI, Switzerland}

\author{E. Skoropata}
\affiliation{PSI Center for Photon Sciences, CH-5232 Villigen-PSI, Switzerland}

\author{K. Rolfs}
\affiliation{PSI Center for Neutron and Muon Sciences, CH-5232 Villigen-PSI, 
Switzerland}

\author{T. Shang}
\affiliation{PSI Center for Neutron and Muon Sciences, CH-5232 Villigen-PSI, 
Switzerland}
\affiliation{Key Laboratory of Polar Materials and Devices (MOE), School of 
Physics and Electronic Science, East China Normal University, Shanghai 
200241, China}

\author{J. Lyu}
\affiliation{PSI Center for Neutron and Muon Sciences, CH-5232 Villigen-PSI, 
Switzerland}
\affiliation{CAS Key Laboratory of Magnetic Materials and Devices, Ningbo 
Institute of Materials Technology and Engineering, Chinese Academy of Sciences, 
Ningbo 315201, China}

\author{E. Pomjakushina}
\affiliation{PSI Center for Neutron and Muon Sciences, CH-5232 Villigen-PSI, 
Switzerland}

\author{M. Medarde}
\affiliation{PSI Center for Neutron and Muon Sciences, CH-5232 Villigen-PSI, 
Switzerland}

\author{H. M. R{\o}nnow}
\affiliation{Institute of Physics, Ecole Polytechnique F\'ed\'erale de 
Lausanne (EPFL), CH-1015 Lausanne, Switzerland}

\author{B. Normand}
\affiliation{PSI Center for Scientific Computing, Theory and Data, 
CH-5232 Villigen-PSI, Switzerland}
\affiliation{Institute of Physics, Ecole Polytechnique F\'ed\'erale de 
Lausanne (EPFL), CH-1015 Lausanne, Switzerland}

\author{M. Radovic}
\affiliation{PSI Center for Photon Sciences, CH-5232 Villigen-PSI, Switzerland}

\author{T. Schmitt}
\email[]{thorsten.schmitt@psi.ch}
\affiliation{PSI Center for Photon Sciences, CH-5232 Villigen-PSI, Switzerland}

\date{\today}

\begin{abstract}
We probe the spectrum of elementary excitations in \SIO~by using 
heterostructured [(\SIO)$_m$/(SrTiO$_3$)$_l$] samples to approach the bulk 
limit. Our resonant inelastic x-ray scattering (RIXS) measurements at the 
Ir $L_3$-edge reveal a robust low-lying collective magnetic mode with an
antiferromagnetic (AF) dispersion similar to the insulators \onelayer~and 
\twolayer, albeit with a large gap and much larger linewidth. At higher 
energies we find the spin-orbit exciton, also strongly broadened, but with 
an inverted dispersion and doubled periodicity that are controlled by the 
charge hopping. These results demonstrate that the AF paramagnon persists, 
somewhat counterintuitively, far into the metallic regime of the insulator-metal
transition driven by the degree of confinement in the heterostructure. We 
conclude that these two excitations, which are contrasting but coexisting 
hallmarks of strong AF pseudospin and charge fluctuations in a 
spin-orbit-coupled Mott-Slater material, are properties intrinsic to 
the ground state of semimetallic perovskite \SIO.
\end{abstract}

\maketitle

Iridium oxides of the Ruddlesden-Popper series Sr$_{n+1}$Ir$_n$O$_{3n+1}$ 
have attracted extensive interest in the search for unconventional phenomena 
arising from the interplay between strong electronic correlations and large 
spin-orbit coupling (SOC) \cite{Kim2008,Moon2008,Watanabe2010,Martins2011,
Arita2012}. Although the compounds \onelayer~and \twolayer, which exhibit 
collective charge localization and antiferromagnetic (AF) ordering of the 
spin-orbit-entangled $J = 1/2$ pseudospins, have been classified as spin-orbit 
Mott insulators (MIs), their energy gap is one order of magnitude smaller 
than that of a typical 3$d$ MI. Indeed the end-member of the series, \SIO, 
appears to be a nonmagnetic, multiband semimetal \cite{Liu2016a,Nie2015a, 
Fujioka2017}, whose remarkably narrow low-lying electronic bands and 
significant admixture of $J = 1/2$ and $3/2$ states indicate that these 
materials lie in the intermediate-coupling, or ``Mott-Slater'' regime
\cite{Hao2019,Yang2020}. Further, the prediction that SOC makes \SIO~a Dirac 
semimetal \cite{Zeb} places it at an exciting nexus of condensed matter 
physics where correlated quantum materials with competing spin, charge, 
and lattice \cite{Nie2015a} energy scales converge with topological phenomena 
\cite{Chen2015}.

\begin{figure}[t]
\includegraphics[width=\linewidth]{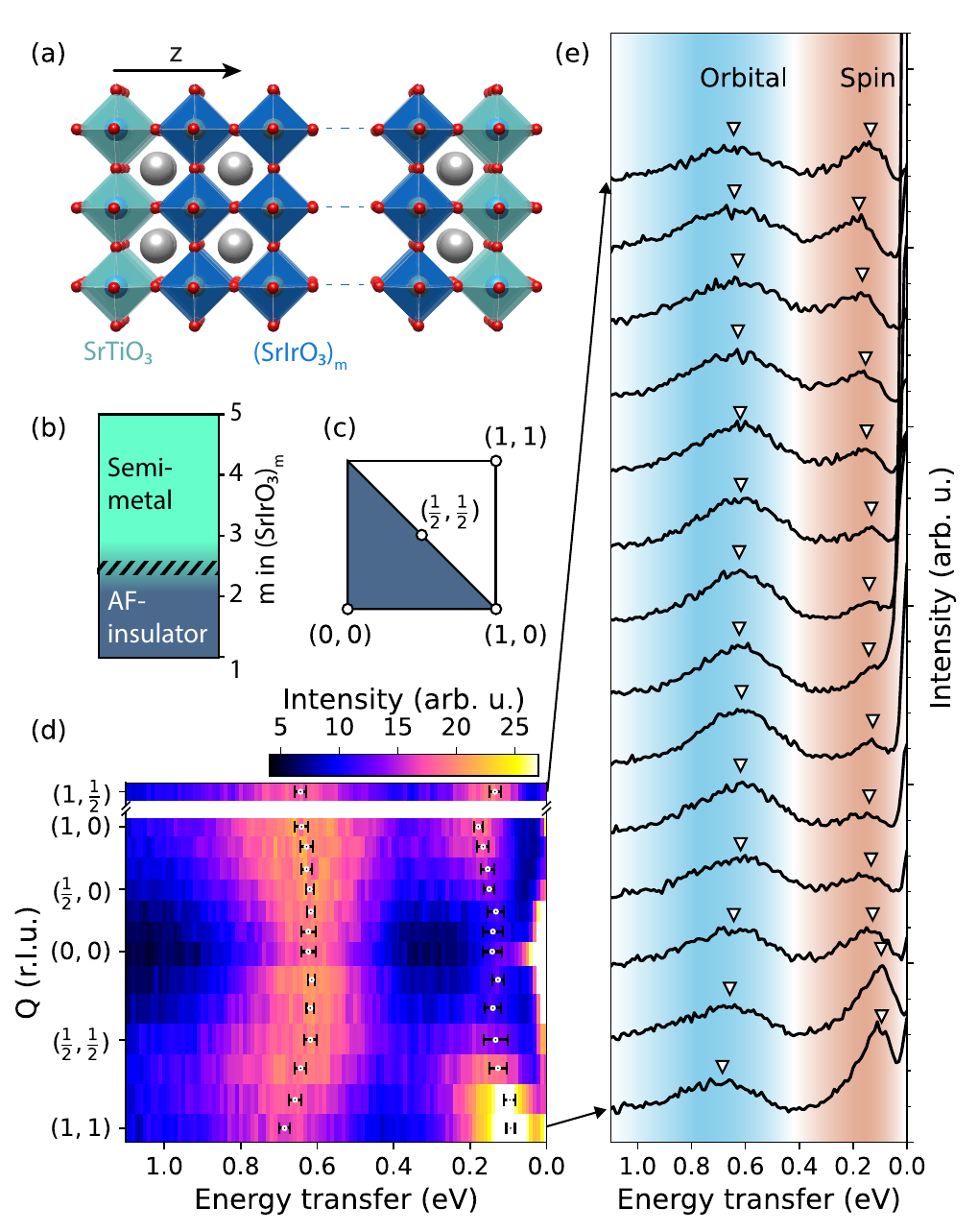}
\caption{{\bf RIXS investigation of (\SIO)$_m$(SrTiO$_3$)$_l$ 
heterostructures.} (a) Representation of (\SIO)$_m$(SrTiO$_3$)$_1$ 
heterostructures, which were grown on a (001)-oriented SrTiO$_3$ 
single-crystal substrate. The samples investigated here had $m = 3$, 
4, and 7 with $l = 1$ and $m = 10$ with $l = 4$. 
(b) Phase diagram of the (\SIO)$_m$(SrTiO$_3$)$_l$ system, illustrating 
the bandwidth-controlled MIT \cite{Kong2022} that takes place as the 
confinement is increased from $m = 3$ to 2 \cite{Matsuno2015a}.
(c) Reciprocal space of the pseudo-cubic AF unit cell.
(d) Compilation of Ir $L_3$-edge RIXS spectra at 
all measured ${\vec Q}$ points for the $m = 4$ sample at $T = 20$ K.
(e) Intensity data of panel (d) shown as a function of energy transfer 
for every ${\vec Q}$ point. We identify a low-energy spin excitation 
(light red) and a spin-orbit exciton at intermediate energies 
(light blue).} 
\label{fig1n}
\end{figure}

Nevertheless, the true properties of perovskite \SIO~have remained enigmatic 
due to the unavailability of single-crystal samples. The SOC-induced coexistence 
of heavy and light carriers \cite{Nie2015a,Sen2020} mixes the transport and 
magnetic properties to produce strong magnetoresistance \cite{Fujioka2017,
Hao2019} and spin-Hall effects \cite{Patri2018}, which have potential for 
low-power spintronic applications \cite{PhysRevMaterials.3.051201, Nan16186, 
Liu2019,Yang2023}. The observation of non-Fermi-liquid transport characteristics
has been interpreted as a hallmark of proximity to a metal-insulator transition 
(MIT) \cite{Moon2008,Groenendijk2017,Sen2020}. Contrary to expectations based 
on the $4d$ analog, SrRuO$_3$, the insulating phase proximate to \SIO~appears 
to be AF, and has been reached by the application of chemical pressure 
\cite{Zheng2016,Cui2016}. The extreme interest in unravelling the complex 
electronic and magnetic structure of \SIO~has led a number of authors to 
investigate dimensionality effects using [(\SIO)$_m$/(SrTiO$_3$)$_l$] 
heterostructures [Fig.~\ref{fig1n}(a)]. These studies found that the 
confinement-driven MIT between $m = 2$ and 3 heterostructures 
[Fig.~\ref{fig1n}(b)] reproduces the $n$-driven MIT of bulk 
Sr$_{n+1}$Ir$_n$O$_{3n+1}$ \cite{Matsuno2015a,Schuetz2017,
Groenendijk2017}, but left unexplored the spectrum of spin, charge, 
and orbital excitations on the metallic side. 

In this Letter we use resonant inelastic x-ray scattering (RIXS) to analyze 
the low-lying collective excitations of \SIO/\STO~heterostructures spanning 
the metallic regime beyond the MIT. In every case from $m = 3$ (significantly 
confined) to $m = 10$ (the putative bulk analog), we find a dispersive 
magnetic excitation similar to the pseudospin-wave modes of single-crystalline 
\onelayer~and \twolayer, except in that it has a large gap and up to double 
the linewidth. Counterintuitively, we also find in every case that the $d$-$d$ 
transitions form a clear, if broad, spin-orbit exciton (SOE) whose dispersion 
is dominated by metallic charge motion. Both modes are largely insensitive to 
the degree of confinement, and hence we deduce that strong and short-ranged 
AF pseudospin correlations critically damped by metallic charge fluctuations 
are the intrinsic physics of the multiband, Mott-Slater semimetallic phase of 
perovskite \SIO. 

To study the excitations on the metallic side of the confinement-driven MIT, 
we prepared three high-repetition superlattice heterostructures of the form 
[(\SIO)$_m$(SrTiO$_3$)]$\times$70, with $m = 3$, 4, and 7, and one of the 
form [(\SIO)$_{10}$(SrTiO$_3$)$_4$]$\times$8. All four samples were grown 
epitaxially on a single-crystal SrTiO$_3$ (001) substrate using pulsed laser 
deposition and their layer structure is represented schematically in 
Fig.~\ref{fig1n}(a). Sample growth and characterization details are presented 
in Sec.~S1 of the Supplemental Materials (SM) \cite{sm}. 

RIXS is a method of choice for probing magnetic fluctuations 
with 100 meV energy scales in strongly correlated materials \cite{LeTacon2011,
LeTacon2013,johnny,Das2018,Lu_Science_2021}, and is unique in providing 
momentum-resolved information for thin-film systems. RIXS has been used to 
investigate the perovskite iridates in their bulk \cite{Kim2012,Kim2012b,
MorettiSala2015,Lu2017,clancy2019} and thin-film forms \cite{Lupascu2014,
Paris2020}, and for studies of the two insulating [(\SIO)$_m$(SrTiO$_3$)$_1$] 
heterostructures, namely $m = 1$ and 2 \cite{Meyers2019,Yang2022}. 

Our Ir $L_3$-edge (11.2145 keV) RIXS experiments were performed at 
the ID20 beamline \cite{Sala2013High} of the European Synchrotron Radiation 
Facility (ESRF). We varied the momentum transfer to cover the ($Q_x,Q_y$) 
plane [Fig.~\ref{fig1n}(c)], with the $z$-axis component fixed to $Q_z
 = 3.67 \pm 0.01$ r.l.u.~(indexed to the unit cell of SrTiO$_3$). The 
sample temperature was set to 20 K. 

The RIXS spectra of all four systems exhibited the same overall features, 
and we use the example of the $m = 4$ heterostructure, shown in 
Figs.~\ref{fig1n}(d-e), to summarize these. All spectra have an intense peak 
near zero energy whose fit (to elastic scattering processes and a weak phonon 
contribution) is discussed in Sec.~S2 of the SM \cite{sm}. In the energy-loss 
region 50--300 meV, the spectra exhibit a dispersive excitation with energy 
minima at the zone center and (1,1) points, and maxima on the boundary of 
the putative AF Brillouin zone, whose energy is qualitatively 
similar to the pseudospin-wave modes reported in Sr$_2$IrO$_4$ and 
Sr$_3$Ir$_2$O$_7$ \cite{Kim2012,Kim2012b}. In the range 0.5--0.8 eV, the 
spectra show a further broad and dispersive peak that is common to many Ir 
oxides and has its origin in $d$-$d$ orbital excitations \cite{Kim2012,
Gretarsson2013Crystal-field,Kim2014Excitonic}.

\begin{figure}[t]
\includegraphics[width=\columnwidth]{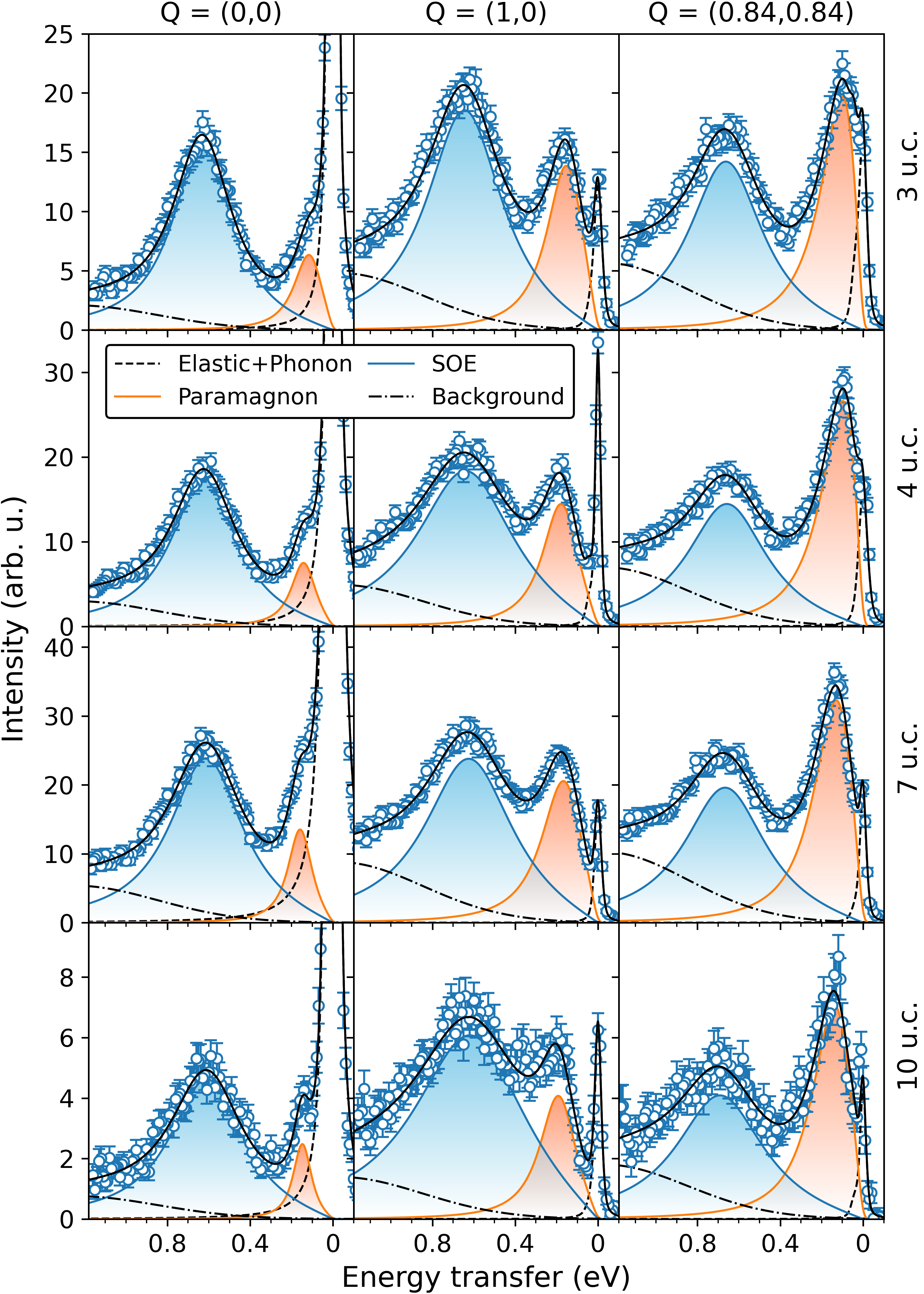}
\caption{{\bf RIXS spectra.} Ir $L_3$-edge spectra at selected momenta 
shown for our four samples. Measured intensity data (symbols) presented 
together with a best fit (black line) composed of elastic and phonon (dashed), 
paramagnon (light red), spin-orbit exciton (SOE, light blue), and high-energy 
contributions (dot-dashed), as detailed in Sec.~S2 of the SM \cite{sm}.}
\label{fig2n}
\end{figure}

In Fig.~\ref{fig2n} we present a quantitative analysis of the RIXS spectra 
at three different ${\vec Q}$ values for all four heterostructures in our 
series, stacked to illustrate their approach to the bulk limit. Definitely 
our most striking qualitative result is that both the pseudospin wave and 
the collective orbital mode are so similar in all respects, despite the 
heterostructures varying from a ``minimal metallic'' configuration ($m = 3$ 
has only one ``bulk'' IrO$_2$ plane and one pair of interfacial 
planes) to one that should truly emulate the bulk. Perhaps the next most 
obvious feature is that the excitations of every heterostructure at every 
${\vec Q}$ are extremely broad (Fig.~\ref{fig2n}), indicating intrinsically 
ultrashort-range spin and orbital correlations. 

\begin{figure}[t]
\includegraphics[width=\columnwidth]{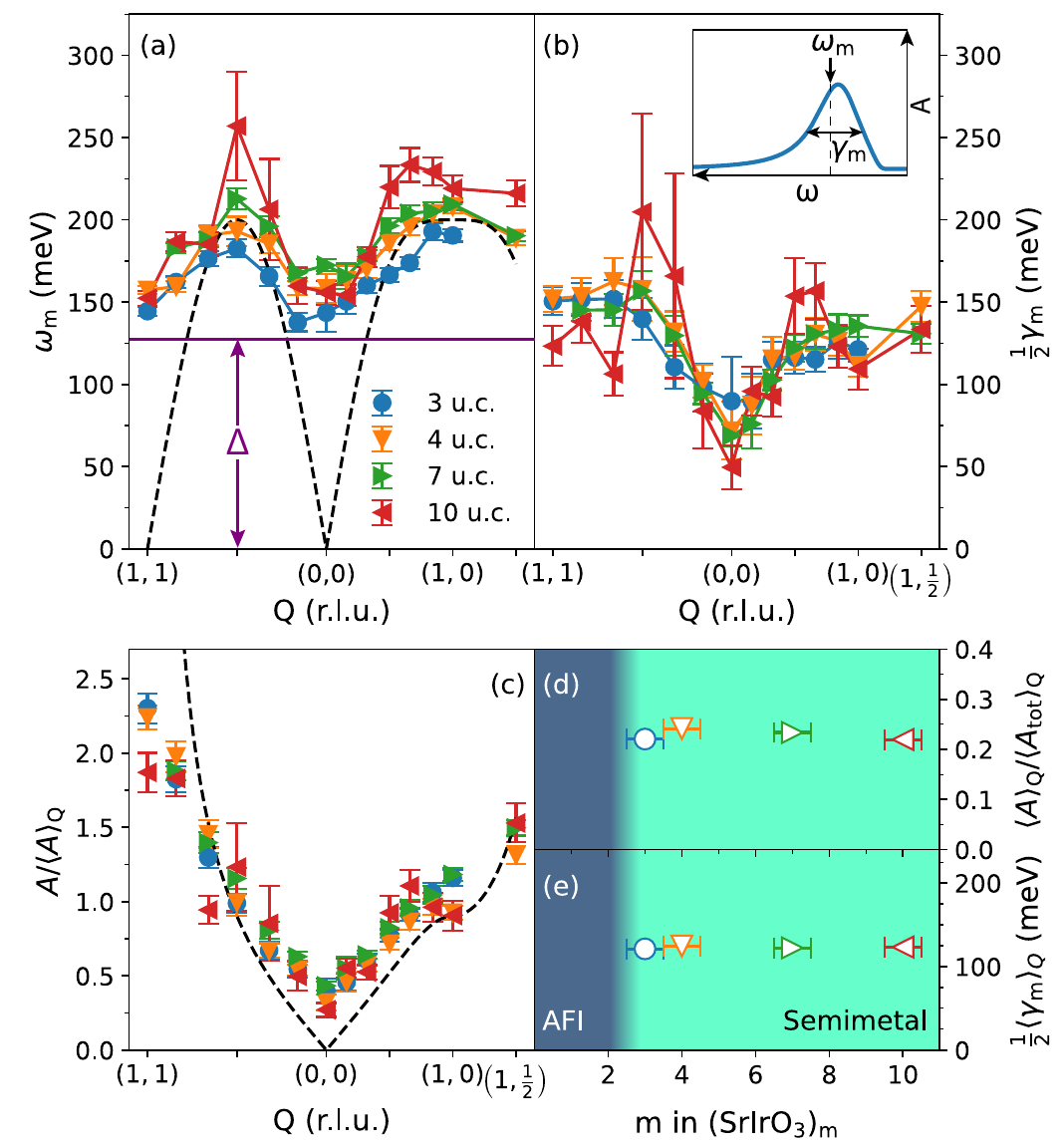}
\caption{{\bf Paramagnon dispersion, linewidth, and intensity.} (a) 
Momentum-dependence of the energy, $\omega_{\rm m} 
({\vec Q})$, of the collective magnetic mode, extracted by fitting to the 
DHO model, for all four samples. The dashed black line shows the dispersion 
of the square-lattice Heisenberg AF for $J = 100$ meV and the 
solid purple line the finite-size gap, $\Delta$, arising due to the ultrashort 
correlation length, $\xi/a \approx 1$. 
(b) Momentum-dependence of the corresponding paramagnon linewidths,
$\gamma_{\rm m} ({\vec Q})/2$, and (c) of the paramagnon spectral 
weights, $A ({\vec Q})$, normalized to the ${\vec Q}$-averaged magnetic 
weight. The black dashed line in (c) is the result for the square-lattice 
Heisenberg AF scaled by an arbitrary constant.
(d) ${\vec Q}$-averaged magnetic weight normalized by the total spectral 
weight and shown as a function of the SrIrO$_3$ layer thickness, $m$. 
(e) Paramagnon linewidth, $\gamma_{\rm m} ({\vec Q})/2$, averaged over 
${\vec Q}$ and shown as a function of $m$.}
\label{fig3n}
\end{figure}

In detail, we performed a multi-component fit of each spectrum to separate 
the contributions of the near-elastic processes, pseudospin and orbital 
excitations, and a high-energy contribution (Fig.~\ref{fig2n}), as described 
in Sec.~S2 of the SM \cite{sm}. We modelled the broad excitations with the 
damped harmonic oscillator (DHO) profile used to describe the strongly 
damped paramagnons observed in a number of doped cuprates 
\cite{lamsal,Robarts} and iron-based superconductors \cite{johnny,Rahn,
Garcia,Lu2022}. The key property of this form [inset Fig.~\ref{fig3n}(b) and 
Sec.~S2] is that it can be used equally when the linewidth, $\gamma_i ({\vec 
Q})/2$, exceeds the bare propagation frequency, $\omega_i ({\vec Q})$, a 
regime we approach in parts of the Brillouin zone (below). As Fig.~\ref{fig2n} 
makes clear, this functional form provides a good fit to the measured line 
shapes of both excitations for all samples at all momenta.

In Fig.~\ref{fig3n} we show the dispersions, linewidths, and normalized 
intensities that we extract for the paramagnon in each of our samples. First, 
the dispersion [Fig.~\ref{fig3n}(a)] is very similar in all four cases. It has 
a generic form familiar from that of magnons arising from nearest-neighbor 
AF interactions with a maximum around 200 meV along the zone boundary 
(black dashed line), except with a huge gap, $\Delta$, of over 130 meV at the 
zone center and a giant damping that is a large fraction of the band energy 
[Fig.~\ref{fig3n}(b)]. Second, this dispersion is similar to that reported 
for the $m = 1$ and 2 heterostructures \cite{Meyers2019,Yang2022}, which 
are thought to be strongly confined and insulating systems. Because the 
confinement-driven MIT occurs at $2 < m < 3$ [Fig.~\ref{fig1n}(b)], it is very 
surprising to find that our samples all exhibit near-identical RIXS spectra 
despite being paramagnetic semimetals. In particular, the bulk-type transport 
and magnetic properties of $m = 10$ samples \cite{Matsuno2015a,Schuetz2017} 
attest that these are well in the semimetallic phase. Thus we deduce that this 
gapped and strongly damped AF pseudospin wave is not related to confinement 
effects in a near-ordered insulator, but is rather an intrinsic property of the 
bulk semimetal.

We interpret the situation as follows. Although the modes of the $m = 1$ 
and 2 heterostructures have been associated with the spin waves of bulk 
\onelayer~and \twolayer~\cite{Meyers2019}, the damping is much larger at all 
momenta and the large gap in all the heterostructures, metallic or insulating, 
well exceeds the values obtained in either material. In fact the two are quite 
different, with \onelayer~behaving as a true MI and square-lattice Heisenberg 
AF, showing a large charge gap and minimal spin gap \cite{Calder2018,
Bertinshaw2020}, whereas \twolayer~has an interlayer spin gap \cite{Kim2012} 
and a charge gap \cite{Mazzone2022} of the same order as the magnetic energy 
scale \cite{MorettiSala2015,Lu2017}, which is a characteristic of the Mott-Slater 
regime \cite{Hao2019,Yang2020}. The giant paramagnon linewidth observed in 
both insulating \cite{Meyers2019} and metallic heterostructures 
[Fig.~\ref{fig3n}(b)], $\gamma_{\rm m} \approx \omega_{\rm m}$, is a hallmark 
of near-unitary scattering, which limits the lifetime $\tau = 1/\gamma_{\rm m}$, 
and hence the spin correlation length ($\xi \approx a$), to the nearest-neighbor 
(n.n.) level, while the opening of the huge gap, $\Delta \approx J (a/\xi)$, in 
Fig.~\ref{fig3n}(a) is a finite-size effect stemming from the correspondingly 
tiny correlated magnetic regions.

To characterize this strong scattering, we first note that the 
${\vec Q}$-dependence of the peak intensity is also nearly identical for all 
four samples [Fig.~\ref{fig3n}(c)], with its maximum at ${\vec Q} = (1,1)$ 
and minimum at ${\vec Q} = (0,0)$ again characteristic of a square-lattice 
Heisenberg AF \cite{Coldea2001}. Next we observe that the dispersions in 
Fig.~\ref{fig3n}(a) show only minor bandwidth changes, falling from around 
80 meV in the bulk limit ($m = 10$) to 60 meV in our more confined samples. 
The slight increase in the zone-boundary energy as the electronic system is 
deconfined suggests the role of interactions with the charge carriers.

Before pursuing this, we finish our discussion of the paramagnon linewidth 
shown in Fig.~\ref{fig3n}(b). This varies by a factor of 2 from the zone 
center to the zone edge, but its ${\vec Q}$-average changes remarkably 
little with the degree of confinement across the heterostructure series, 
as we show in Fig.~\ref{fig3n}(e). While hybridization with the TiO$_2$ 
layers and effective doping of the interfacial IrO$_2$ layers cannot be excluded 
as factors contributing to the strong damping, at minimum our $m = 10$ sample 
has rather few interfacial layers and thus one expects that the right-hand side of 
Fig.~\ref{fig3n}(e) provides a good estimate of the intrinsic damping in bulk 
\SIO. As a further probe of possible interfacial effects, in Fig.~\ref{fig3n}(d) 
we show the average magnetic intensity normalized to the total spectral 
intensity, and again find no meaningful change. This result constitutes one 
of our most important indicators that all the features we observe are intrinsic 
to the \SIO~system, as distinct modes of the interfacial IrO$_2$ layers would 
be expected to scale as $1/m$. 

\begin{figure}[t]
\includegraphics[width=\columnwidth]{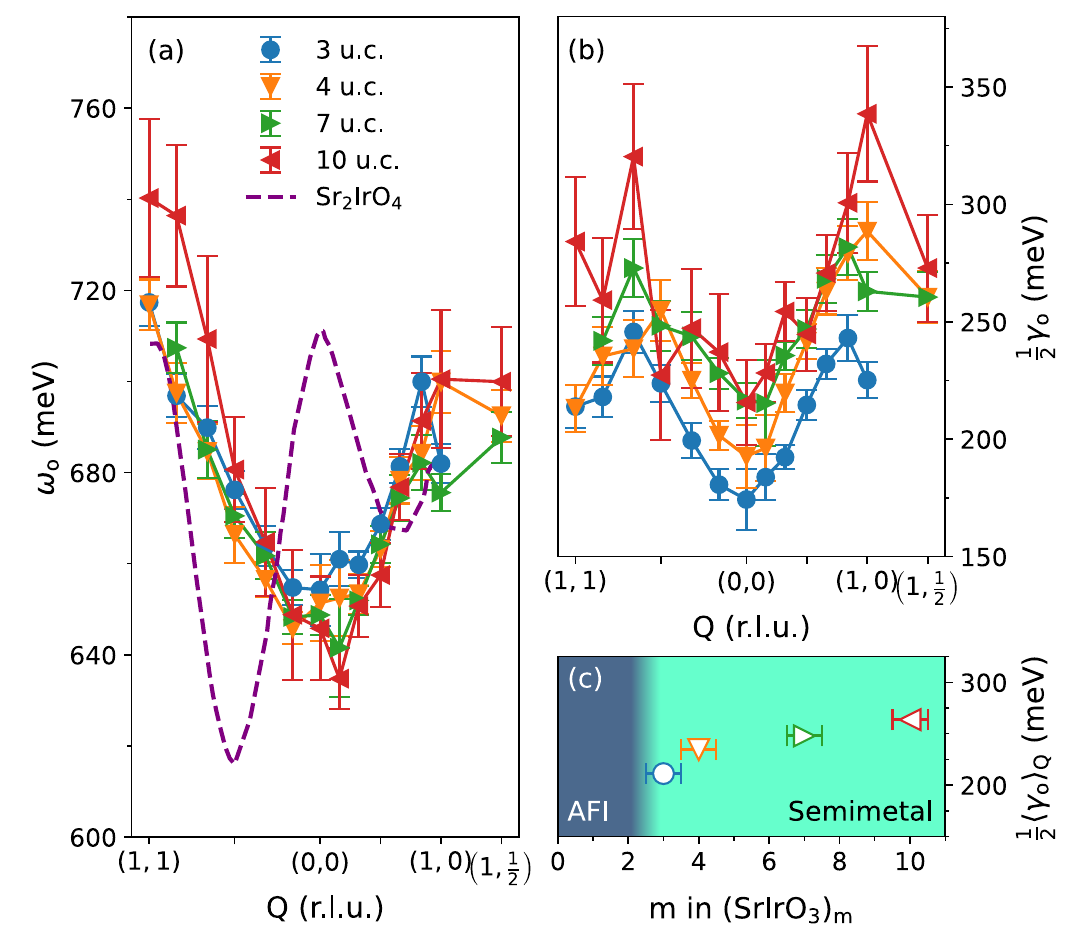}
\caption{{\bf Spin-orbital exciton.} (a) Momentum-dependence of the SOE 
excitation energy, $\omega_{\rm o} ({\vec Q})$, extracted from the DHO fit for 
all four samples. The purple line shows the SOE dispersion in Sr$_2$IrO$_4$, 
taken from the data of Ref.~\cite{Kim2012} as described in Sec.~S3 of the SM 
\cite{sm}. (b) Momentum-dependence of the corresponding linewidth, 
$\gamma_{o} ({\vec Q})/2$. (c) $\gamma_{\rm o} ({\vec Q})/2$ averaged 
over ${\vec Q}$ and shown as a function of $m$.}
\label{fig4n}
\end{figure}

Turning now to the orbital excitation, we apply the same DHO analysis to 
extract the energy and linewidth. Figure \ref{fig4n}(a)
shows that this excitation forms a broad and dispersive mode with a bandwidth 
of 100 meV, conventionally described as a ``spin-orbit exciton'' (SOE)
\cite{Kim2012,Kim2014Excitonic,Gretarsson2019,Zimmermann2023,Wang2024}. 
In Fig.~\ref{fig4n}(a) it is difficult to discern much change in its dispersion 
caused by the confinement. As one might anticipate from Fig.~\ref{fig2n}, the 
linewidth of the SOE is very high, and Fig.~\ref{fig4n}(b) shows that, like the 
damping of the paramagnon [Fig.~\ref{fig3n}(b)] it also varies by 80-120 meV 
as a function of ${\vec Q}$. In this case we note a systematic trend towards a 
larger linewidth as the confinement is removed, quantified in 
Fig.~\ref{fig4n}(c), which contrasts with the behavior of the paramagnon and 
presumably reflects the progressive delocalization of electrons contributing to 
the relevant scattering processes. 

Returning to Fig.~\ref{fig4n}(a), the dispersion of the SOE is completely 
different from that of \onelayer~\cite{Kim2012,Kim2014Excitonic}, whose form 
is characteristic of a magnetic insulator. Instead, the upward dispersion around 
${\vec Q} = (0,0)$ follows the form found in the paramagnetic metal 
Sr$_2$RhO$_4$ \cite{Zimmermann2023} and the doubled ${\vec Q}$-space 
periodicity relative to \onelayer~[purple line in Fig.~\ref{fig4n}(a)] shows no 
evidence of AF character. From thius result we conclude that, despite the 
short-range-correlated magnetic background in \SIO~systems, the propagation of 
the SOE is dominated by metallic hopping processes of the conduction electrons. 

The origin of such strongly contrasting SOE behavior is that the physics of the 
AF insulator is governed by the narrow, filled lower Hubbard-type band formed 
from the $J = 1/2$ electronic states, whereas in a metallic system it is controlled 
by charge hopping in the $J = 3/2$ band \cite{Zimmermann2023}. The surprise 
in our results is that a paramagnon so similar to the AF insulator coexists with
a manifestly metallic SOE. A clue to this coexistence is provided by the behavior 
of both hole- \cite{clancy2019} and electron-doped \cite{Gretarsson2016}
\onelayer, where the paramagnon and the constraining effect of the magnetic 
background on the mobile holes persists out to dopings in excess of 10\%. Thus 
the paramagnon can retain its definition also when ``self-doping'' of the upper 
$J = 1/2$ band occurs as the $J = 3/2$ band becomes broad enough to make 
the system metallic. The finite concentration of carriers causing the bad-metal 
transport characteristics, whose correlation-controlled charge fluctuations 
cause the giant linewidths, also decouples the behavior of the SOE from the 
physics of the narrow $J = 1/2$ band. The fact that the SOEs of both 
heterostructured and bulk \SIO~systems can be so different from \onelayer, 
while the magnetic dispersion has so many similarities, underlines the complex 
interplay of structure and correlations in this class of materials \cite{Liu2021}. 

A further important argument that the giant paramagnon damping is an intrinsic 
property of \SIO~can be found in the extensive reports of non-Fermi-liquid and 
quantum critical behavior in this material \cite{Cao2007,Moon2008,Sen2020}. 
\SIO~heterostructures exhibit the fingerprints of both Mott physics, in the form
of the bad-metal (or strange-metal) resistivities \cite{Matsuno2015a}, and 
Slater physics, in the form of the strong AF spin fluctuations we observe 
despite the charge being delocalized. Indeed the emerging theoretical consensus 
\cite{Hao2019,Yang2020} places \SIO~in the Mott-Slater regime of intermediate 
correlations. Our observation of near-critically damped magnetic and orbital 
fluctuations ($\gamma_{\rm i} \approx \omega_{\rm i}$) indicates that their 
mean free path is of order one lattice constant ($\xi/a \approx 1$), which 
stands to reason if the Mott-Slater charge fluctuations drive the system 
towards the Mott-Ioffe-Regel limit. This very-bad-metal behavior of the 
quasiparticle scattering in \SIO~has also been shown \cite{Sen2020} to 
approach the Planckian limit \cite{Grissonanche2021} at finite temperatures. 

Finally, the most important issue when drawing our conclusions from 
heterostructures concerns the effect of the TiO$_2$ layers. It is known that 
Ti-O-Ir hybridization occurs \cite{Matsuno2015a,Kim2016}, that the AF ground 
state at $m = 1$ changes with the SrTiO$_3$ thickness ($l > 1$) \cite{Hao2017}, 
and that the $m = 1$ magnon dispersion changes slightly when using CaTiO$_3$ 
\cite{Yang2022}. The rotation and distortion of IrO$_6$ octahedra caused by 
the lattice mismatch are also found to affect the interfacial layers 
\cite{Nie2015a,PhysRevLett.119.077201,Meyers2019}. To judge whether the 
properties we observe could be a consequence of heterostructuring, rather than 
being intrinsic to \SIO, we have argued that they should show a systematic 
dependence on the layer thickness. This was not the case for the paramagnon 
broadening in Fig.~\ref{fig3n}(e) or the SOE broadening in Fig.~\ref{fig4n}(c). 
More specifically, if a mode were interfacial, its spectral weight should be 
proportional to the interface/bulk ratio of the different samples. As we showed 
in Fig.~\ref{fig3n}(d), the intensity of the paramagnon is a rather steady 
22-24\% of the total weight, even as the nominal ratio of interfacial IrO$_2$ 
layers changes from 67\% to 20\%. To complete this argument, we probed the 
local chemical environment of the interfaces by Ti $L$-edge X-ray absorption 
spectroscopy (XAS), as we detail in Sec.~S1B of the SM \cite{sm}. This 
confirmed first that finite Ti-O-Ir hybridization takes place and second that 
the XAS intensities from Ti atoms in different environments do scale linearly 
with the nominal ratio given by the heterostructure geometry. Hence the 
magnetic mode we observe must be a property of the bulk, while the 
hybridization remains a minor quantitative effect. 

In summary, we have performed RIXS measurements to unveil the spectrum 
of elementary excitations in heterostructured \SIO. Investigating a series of 
systems on the semimetallic side of the confinement-driven MIT confirms the 
presence in all spectra of two universal features that are also shared with the 
AF insulator. At low energy is a broad and gapped pseudospin paramagnon 
that reveals the simultaneous presence of strong AF correlations and strong 
quasiparticle scattering. In counterpoint is a spin-orbit exciton (SOE) at 
higher energy, whose dispersion and extreme linewidth reflect only the rapid 
charge fluctuations of semimetallic \SIO, and not the strong AF character. 
However, charge fluctuations arising from the non-Fermi-liquid nature of the 
strange-metal phase in \SIO~cause the magnetic fluctuations to remain 
ultrashort-ranged, and hence provide additional evidence for a quasiparticle 
scattering rate near the unitary limit. The remarkable robustness of both 
nominally incompatible excitations, the AF paramagnon as confinement is 
removed and the metallic SOE as confinement is increased, leads us to 
conclude that both are generic properties of bulk \SIO.

{\it Acknowledgments.} We thank V. Favre and X. Lu for helpful contributions. 
This work was supported by the Swiss National Science Foundation through the 
Sinergia network Mott Physics Beyond the Heisenberg Model (MPBH, Research 
Grants CRSII2 160765/1 and CRSII2 141962) and through research project 
No.~200021 178867. We acknowledge the ESRF for the provision of beam-time 
on ID20. The Ti $L$-edge XAS was performed at the ADRESS beamline of the 
Swiss Light Source at the Paul Scherrer Institut (PSI).

\newpage

\onecolumngrid

\noindent
{\large {\bf {Supplementary Material to accompany the article}}}

\vskip4mm

\noindent
{\large {\bf {Coexistence of insulator-like paramagnon and metallic spin-orbit exciton modes in \SIO}}}

\vskip4mm

\noindent
E. Paris, W. Zhang, Y. Tseng, A. Efimenko, C. Sahle, V. N. Strocov, E. Skoropata, K. Rolfs, T. Shang, J. Lyu, E. Pomjakushina, M. Medarde, H. M. Rønnow, B. Normand, M. Radovic, and T. Schmitt

\vskip6mm

\twocolumngrid

\section{Sample synthesis and characterization}

\subsection{Superlattice structure}

Superlattices of the forms [(\SIO)$_m$(SrTiO$_3$)]$\times$70, with $m = 3$, 4, and 7 (henceforth 3 u.c., 4 u.c., and 7 u.c.), and [(\SIO)$_{10}$(SrTiO$_3$)$_4$]$\times$8 (henceforth 10 u.c.) were grown on (001)-oriented SrTiO$_3$ substrates using pulsed-laser deposition (PLD). The polycrystalline SrIrO$_3$ target and a single crystal SrTiO$_3$ target were ablated using a solid-state YAG laser with a fluence of approximately 1 Jcm$^{-2}$. Epitaxial growth conditions were achieved with an oxygen partial pressure of 0.1 mbar while the temperature of the substrate was kept at 690 $^\circ$C. In preparing our 3, 4, and 7 u.c.~samples, we realized a large number of repetitions to benefit from the large penetration depth of 12 keV X-rays in order to achieve a high RIXS signal intensity. In the 10 u.c.~sample, a smaller number of repetitions was sufficient to ensure a reasonable RIXS signal. A capping layer of 5 unit cells of SrTiO$_3$ was deposited on top of each sample to prevent surface degradation. 

\begin{figure}[t]
\centering
\includegraphics[width=\linewidth]{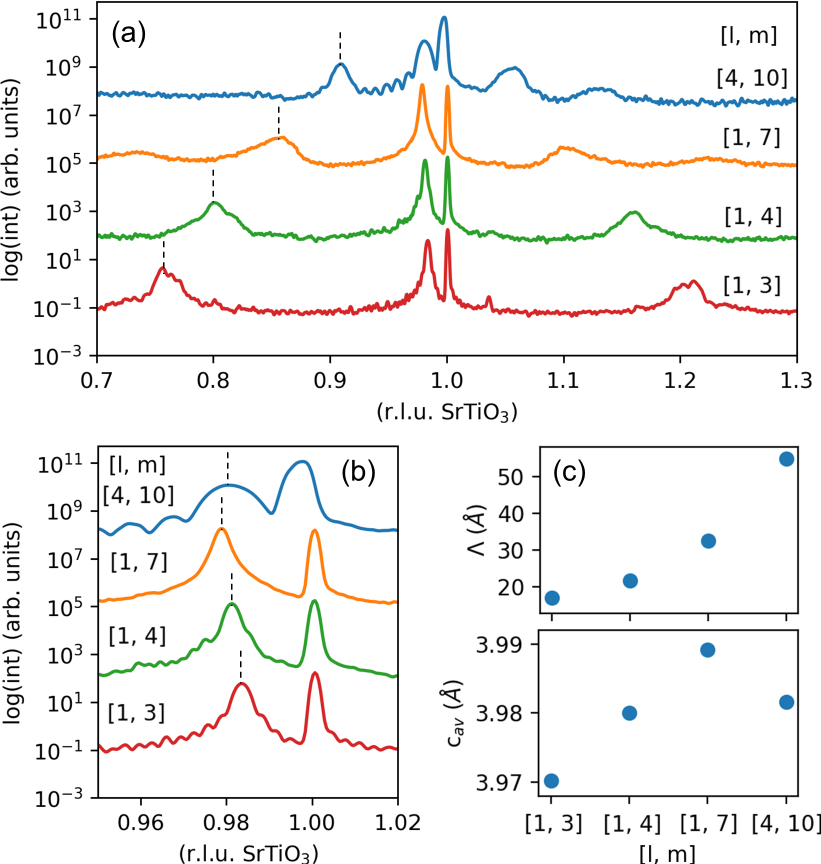}
\caption{$\theta$-$2\theta$ X-ray diffraction (XRD) scans of the $m = 3$, 4, 7, and 10 u.c.~samples around the (002) reflection,  indexed with respect to the SrTiO$_3$ substrate. The change in structure is indicated by the dashed lines in panel (a) marking the superlattice period, $\Lambda$, and in panel (b) marking the average $c$-axis lattice parameter, $c_{\rm{av}}$. $\Lambda$ and  $c_{\rm{av}}$ are shown in panels (c) and (d) respectively.\label{fig:XRD}}
\end{figure}

Figure~\ref{fig:XRD} shows $\theta$-$2\theta$ x-ray diffraction (XRD) scans of all four samples measured using a Bruker D8 Discover with a one-dimensional detector working at wavelength $\lambda = 1.5406$ \AA~(Cu K$_{\alpha}$). The appearance of high-frequency oscillations only in the $m = 10$ data is a consequence of the fact that our $m = 3$, 4 and 7 samples are very thick (respectively 280, 350, and 560 perovskite unit cells, compared to 112 for $m = 10$ and 80 in Ref.~\cite{Matsuno2015a}). The superlattice [Fig.~\ref{fig:XRD}(a)] and film reflections [Fig.~\ref{fig:XRD}(b)] change systematically across the sample series and have separate and sharp peaks, with no evidence for other layerings. From these we obtain the period, $\Lambda = l c_{\rm{STO}} + m c_{\rm{SIO}}$ [Fig.~\ref{fig:XRD}(c)], and the average out-of-plane lattice parameter of the film, $c_{\rm av} = (l c_{\rm{STO}} + m c_{\rm{SIO}})/(l + m)$ [Fig.~\ref{fig:XRD}(d)]. The values of $m$ and $l$ deduced from the measured values of $\Lambda$ and $c_{\rm av}$ are [$l$, $m$] = [1, 3], [1, 4], [1, 7], and [4, 10], with $c_{\rm{STO}} = 3.905$~\AA~and $c_{\rm{SIO}} = 4.005$~\AA. The uncertainty in the layer thickess is estimated at 0.3 u.c. 

\begin{figure}[t]
\centering
\includegraphics[width=\linewidth]{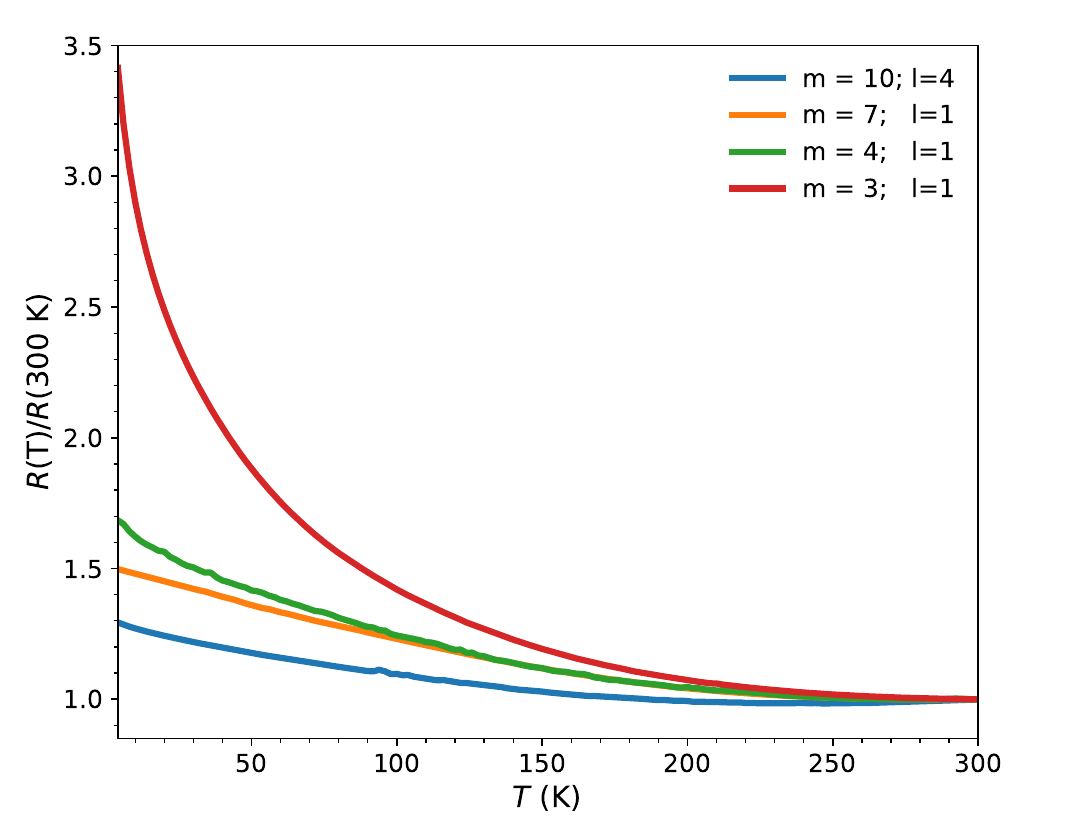}
\caption{Resistivity characteristics of all four heterostructure samples as a function of temperature, measured upon cooling and normalized to their room-temperature values. No significant differences were observed during warming. \label{fig:R}}
\end{figure}

We also characterized all four heterostructures with resistivity measurements performed by the four-point method on a Quantum Design PPMS 9T. As Fig.~\ref{fig:R} shows, in every case we found the resistance upturn at low temperatures characteristic of bad-metal behavior and consistent with the results for the metallic samples of Refs.~\cite{Matsuno2015a,Fan2017,Kleindienst2018}. Although direct comparisons between samples were precluded by widely varying contact resistances, normalization to the 300 K resistance produces a systematic trend towards a stronger upturn as the $m$-induced MIT is approached.

\subsection{X-ray absorption spectroscopy (XAS)}

\begin{figure}[t]
\centering
\includegraphics[width=0.94\linewidth]{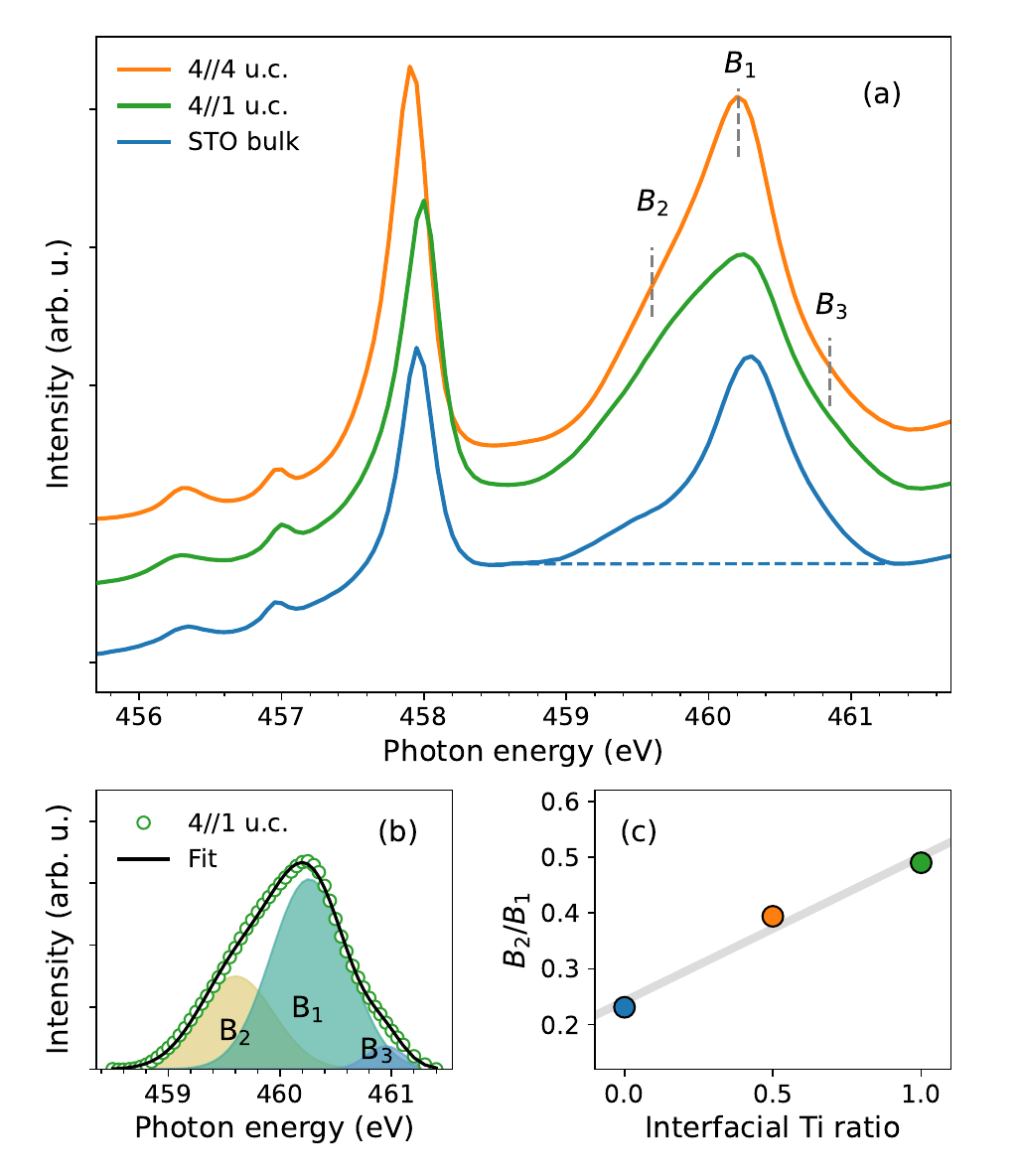}
\caption{\label{fig:xas}(a) Ti $L_3$-edge XAS spectra measured by collecting the total fluorescence yield using linearly polarized light at 20 K. The incidence angle on the sample surface was 15$^\circ$ (10$^\circ$ for the 4//4 u.c.~sample). All spectra were normalized to the total spectral area in the region from 455 to 475 eV.The dashed line represents the background that was subtracted before fitting the separate contributions. (b) Fit of the measured peak with contributions $B_1$, $B_2$, and $B_3$. (c) Ratio $B_2/B_1$ as a function of the fraction of interfacial Ti ions in the sample. }
\end{figure}

To probe the local chemical environment of the IrO$_2$-TiO$_2$ interfaces, we performed Ti L-edge XAS on two $m = 4$ samples and compared the results with a SrTiO$_3$ (STO) reference. These measurements were performed at the ADRESS beamline \cite{Strocov2010High-resolution} of the Swiss Light Source at the Paul Scherrer Institut (PSI). For this measurement we prepared a sample with fewer (40) superlattice repetitions, because of the lower penetration depth of the relevant soft x-rays when compared to hard x-rays used for RIXS. The XAS data are shown in Fig.~\ref{fig:xas}(a), with the standard 4 u.c.~sample ($m = 4$, $l = 1$) shown as the green line and the bare SrTiO$_3$ substrate as the blue line. For comparison we prepared a [(\SIO)$_4$(SrTiO$_3$)$_4$]$\times$8 sample ($l = 4$, labelled ``4//4 u.c.''), which differs from the 4 u.c.~sample by increasing the thickness of the STO layers and hence reducing the density of interfacial Ti atoms. 

When compared to STO, the 4 u.c.~sample (labelled 4//1 u.c. in Fig.~\ref{fig:xas}) clearly has a larger spectral weight in the energy region related to the $2p_{3/2} \rightarrow 3d (e_g)$ transition, labeled $B_2$ in Fig.~\ref{fig:xas}(a).  Such a difference can be due either to the presence of a minority $d^1$ component for the Ti ions in the heterostructure \cite{Jose2019} or to a local structural distortion of the TiO$_6$ octahedra at the interface, which would act to alter the local crystal-field splitting.  The presence of a Ti $d^1$ component would imply hybridization between Ti and Ir ions through the oxygen ligands across the interface, in agreement with theoretical calculations \cite{Matsuno2015a} and optical spectroscopy results on similar heterostructures \cite{Kim2016}. On the other hand, interfacial structural distortions have been reported in similar heterostructures \cite{Meyers2019}. Regardless of the origin of this effect, the change we measure in the XAS spectrum is an indication that the local electronic configuration of the Ti ions at the STO/SIO interface differs from that in bulk STO. 
 
To show that the $B_2$ component in the XAS spectrum is indeed due to the interfaces, we compare the 4 u.c.~and STO samples with the 4//4 u.c.~sample. We first subtract a constant background [an example is shown in Fig.~\ref{fig:xas}(a) for bulk STO] and fit the background-subtracted data with three Gaussian functions, two of which describe the $B_1$ and $B_2$ peaks while the third reproduces an additional peak ($B_3$) at higher energy that is present in all samples. The $B_1$ and $B_2$ Gaussians are constrained to have equal width and an example fit is shown in Fig.~\ref{fig:xas}(b) for 4 u.c. In Fig.~\ref{fig:xas}(c), we report the peak ratio, $B_2/B_1$, as a function of the nominal fraction of interfacial Ti ions for each sample, expressed as $N_{\rm int}/(N_{\rm int} + N_{\rm bulk})$. The peak ratio is $B_2/B_1 = 0.23$ in STO, 0.39 for 4//4 u.c., and 0.49 for 4 u.c. That the ratio for 4//4 u.c. lies half-way between those of 4 u.c.~and STO, within the uncertainty of these measurements, confirms that the $B_2$ signal has its origin in interfacial Ti atoms. Such a linear evolution confirms the hypothesis made in the main text that the interfacial fractions of Ti and Ir in our heterostructures are close to their nominal values, on the basis of which we eliminated interfacial effects in the magnetic signal. 

\section{Fitting RIXS data}

Here we present the strategy and data-processing by which we use the RIXS data to extract the different elementary excitations. It has been shown that in this type of multilayer, the structure of the SrIrO$_3$ layer is tetragonal up to a thickness of 12 unit cells, above which an orthorhombic distortion occurs \cite{Nan16186}. Thus we base our analysis on the reciprocal-space vectors of the tetragonal system, referring to the structure of the STO substrate with $a = b = 3.9$ \AA. We show first the full fitting function we apply at each separate wave vector, {\bf Q}, and then describe its separate components. This function has the form 

\onecolumngrid

\begin{align}
\label{eq:ff}
I(\omega) &= pV(\omega,\Gamma_{\rm e},\eta, I_{\rm e}) \nonumber \\
&\qquad + g_{\text{res}} \circledast \Big[ 
    \big( 
        L(\omega,\omega_{\rm p},\Gamma_{\rm p},I_{\rm p})
        + D(\omega,\omega_{\rm m},\gamma_{\rm m},I_{\rm m})
        + D(\omega,\omega_{\rm o},\gamma_{\rm o},I_{\rm o}) 
    \big) B(\omega,T) \nonumber \\
&\qquad\quad + 
    \big( 
        L(\omega,-\omega_{\rm p},\Gamma_{\rm p},I_{\rm p})
        + D(\omega,-\omega_{\rm m},\gamma_{\rm m},I_{\rm m}) 
    \big) B^*(\omega,T)
\Big] + G(\omega,\omega_{\rm b},\Gamma_{\rm b},I_{\rm b})
\end{align}
where
\begin{align}
\label{eq:ffc}
pV(\omega,\Gamma_{\rm e},I_{\rm e}) &= I_{\rm e} \bigg[ \eta\frac{2\sqrt{\ln2}}{{\sqrt \pi}\Gamma_{\rm e}}\exp\bigg({-4\ln2 \frac{\omega^2}{\Gamma_{\rm e}^2}}\bigg)+(1-\eta)\frac{1}{\pi}\frac{\Gamma_{\rm e}/2}{\omega^2+(\Gamma_{\rm e}/2)^2}\bigg], \\
L(\omega,\omega_{\rm i},\Gamma_{\rm i},I_{\rm i}) &= \frac{I_{\rm i}}{\pi}\frac{\Gamma_{\rm i}/2}{(\omega-\omega_{\rm i})^2+(\Gamma_{\rm i}/2)^2}, \\
D(\omega,\omega_{\rm i},\gamma_{\rm i},I_{\rm i}) &=  \frac{2\omega I_{\rm i} \gamma_{\rm i}  \omega_{\rm i} }{[(\omega^2 - \omega_{\rm i}^2)^2 + (\omega \gamma_{\rm i} )^2]}, \\
B(\omega,T) &= \frac{1}{1 - e^{-\beta\hbar\omega}}, \\
B^*(\omega,T) &= \frac{e^{-\beta\hbar\omega}}{1 - e^{-\beta\hbar\omega}}, \\
G(\omega,\omega_{\rm b},\Gamma_{\rm b},I_{\rm b}) &= I_{\rm b} \frac{2 \sqrt{\ln 2}}{{\sqrt{\pi}\Gamma_b}} \exp \bigg({-4 \ln 2 \frac{(\omega-\omega_{\rm b})^2}{\Gamma_{\rm b}^2}}\bigg).
\end{align}

\twocolumngrid

Starting with the elastic contribution to the measured signal, for each sample at each momentum we fit the elastic line profile using a pseudo-Voigt function, $pV(\omega, ...)$ in Eqs.~\eqref{eq:ff}, in which the normalized Gaussian and Lorentzian functions have the same full width at half maximum (FWHM) height, $\Gamma_{\rm e}$, and a mixing parameter that we fix as $\eta = 0.47$. 

Next, all our spectra contain a low-energy mode that is thought to be phononic in origin \cite{Meyers2019}, and which we describe with the resolution-limited Lorentzian profile $L(\omega, ...)$ in Eqs.~\eqref{eq:ff}. We fixed the phonon energy to the value $\omega_{\rm p} = 40$ meV reported in previous studies \cite{Meyers2019}, and to improve the fitting stability we also fixed its width to $\Gamma_{\rm p} = 0.45 \Gamma_{\rm e}$. The symbol $\circledast$ in Eqs.~\eqref{eq:ff} represents a frequency convolution with a Gaussian of unit magnitude, $g_{\text{res}}$, that ensures the FWHM of the resolution function, $\gamma_{\rm res} = 30$ meV. 

The damped harmonic oscillator profile $D(\omega, ...)$~\cite{lamsal} was used for both the magnon and SOE modes, with $I_{\rm i}$ the oscillator strength, $\omega_{\rm i}$ the bare propagation frequency of the assumed single broad mode, and $\gamma_{\rm i}$ the damping factor. This model describes on an equal footing the regimes where the mode is underdamped ($\omega_{\rm i} > \gamma_{\rm i}/2$), critically damped ($\omega_{\rm i} = \gamma_{\rm i}/2$), or overdamped ($\omega_{\rm i} < \gamma_{\rm i}/2$). The functions $B(\omega,T)$ and $B^*(\omega,T)$ are the respective weighting terms for the Stokes and anti-Stokes excitations, in which $\beta = 1/k_{\rm B}T$ and $T = 20$ K; both $B$ and $B^*$ have a discontinuity at $\omega = 0$ and thus are defined in the range $|\omega| > \delta$, with $\delta = 5$ meV representing a cutoff for the inelastic contributions. 

\begin{figure*}[t]
\centering
\includegraphics[width=\textwidth]{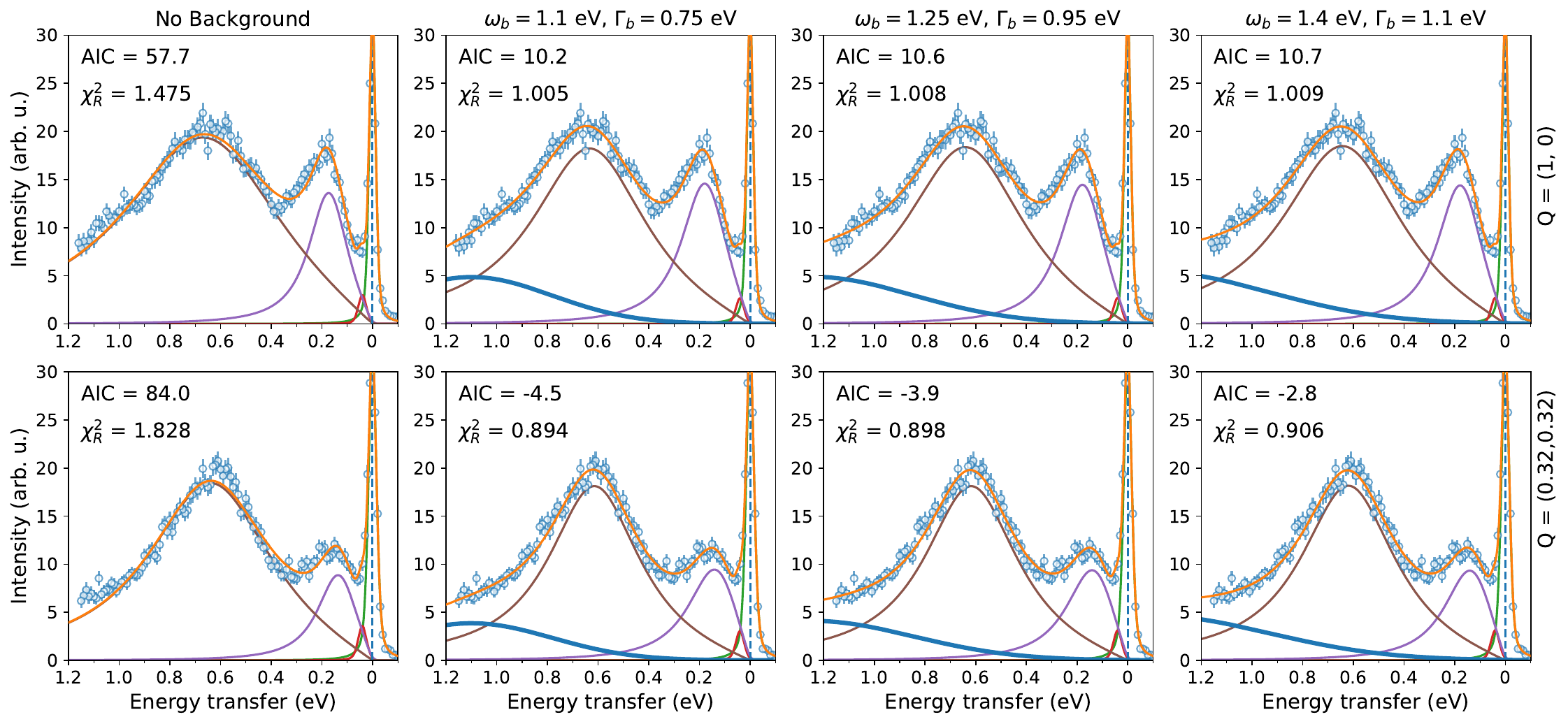}
\caption{\label{ra4} Comparison between intensity fits made using different models for the high-energy background. We compare fits to the intensity measured for one sample (4 u.c.) at two different momenta, ${\bf Q} = (1,0)$ in the upper panels and ${\bf Q} = (0.32,0.32)$ in the lower. The high-energy background, shown by the thick blue line, is modelled as a Gaussian peak with 3 different ($\omega_{\rm b}$,$\Gamma_{\rm b}$) choices, or is set to zero (left panels). The number of free parameters is $N_p = 9$ with no high-energy background and $N_p = 10$ with one. For each fit we specify the values of two fit estimators, $\chi_{\rm R}^2$ and the Akaike information criterion (AIC).  }
\end{figure*}

Finally, all our spectra contain an additional high-energy contribution above 1 eV whose origin lies in excitations beyond the SOE, which are usually expected to be of electronic or electron-hole character. Because this contribution is very broad and has no discernible structure, one fitting strategy is simply to ascribe all of the high-energy signal to the energy, $\omega_{\rm o}$, broadening, $\gamma_{\rm o}$, and intensity, $I_{\rm o}$, of the SOE, choosing to accept uncertainties in these estimated properties rather than to introduce further fitting parameters to cover this range. However, because the intrinsic parameters of the SOE are central to our study, we attempt to model the high-energy background as an additional Gaussian function,  $G(\omega, ...)$ in Eqs.~\eqref{eq:ff},  by adopting the following procedure. We fix the energy ($\omega_{\rm b}$) and the FWHM ($\Gamma_{\rm b}$) of the Gaussian, leaving its overall intensity, $I_{\rm b}$, as the only fitting parameter. To deal with the extra degrees of freedom in choosing $\omega_{\rm b}$ and $\Gamma_{\rm b}$, we computed the momentum-averaged reduced $\chi$-squared ($\langle \chi_{\rm R}^2 \rangle_Q$) of the overall fits to all four of our samples, finding a rather broad and flat minimum in its dependence on the choice of $\omega_{\rm b}$ and $\Gamma_{\rm b}$ within a reasonable range. In Fig.~\ref{ra4}, we select three different ($\omega_{\rm b}$,$\Gamma_{\rm b}$) pairs and show the details of the intensity fit for the illustrative example of the 4 u.c. sample at two different momenta, comparing in the left panels with a fit that assumes no additional high-energy background. 

\begin{figure}[t]
\centering
\includegraphics[width=\linewidth]{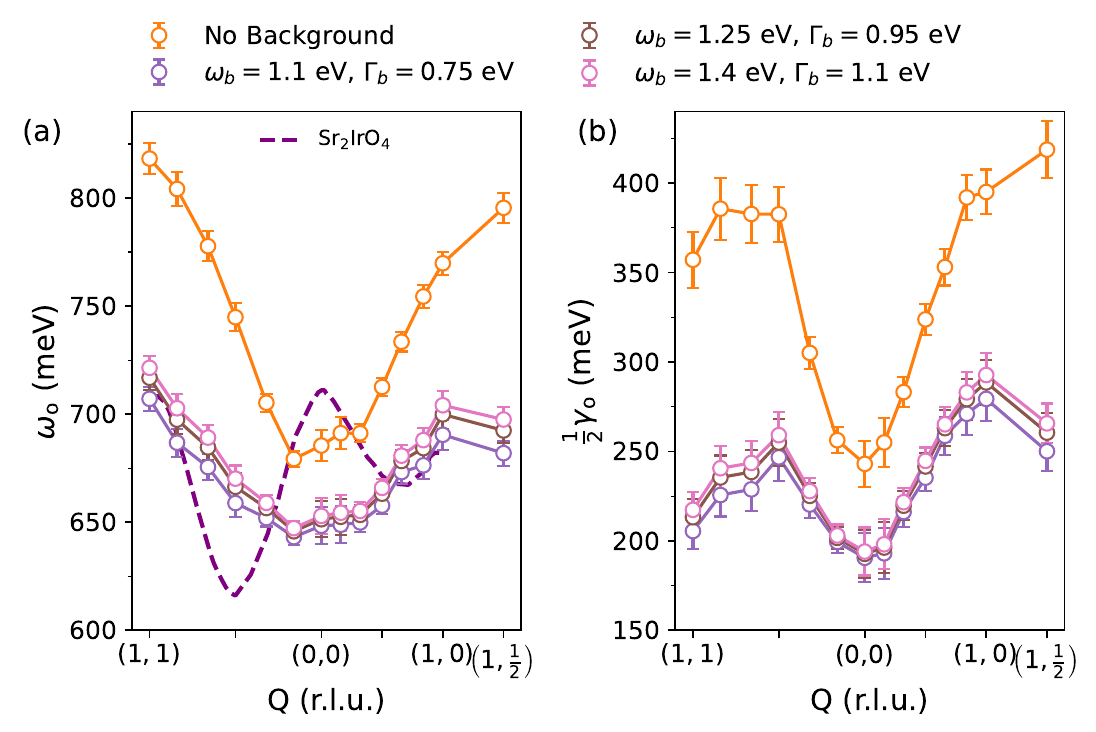}
\caption{\label{ra5} Excitation energy (a) and linewidth (b) of the SOE, shown as functions of the momentum transfer for the 4 u.c. sample when using different fitting models. Shown in orange is a fit with no background, in which all high-energy intensity is ascribed to the SOE. In the other fits we have applied a model with an additional Gaussian peak centered at 1.1 eV with a width of 0.75 eV (purple), at 1.2 eV with a width of 0.95 eV (brown), or at 1.4 eV with a width of 1.1 eV (pink). }
\end{figure}

In order to understand the effects of these background choices on the SOE position and linewidth we extract, in Fig.~\ref{ra5} we compare the four sets of fitting results obtained at all the experimental {\bf Q} values using the three different ($\omega_{\rm b}$,$\Gamma_{\rm b}$) choices shown in Fig.~\ref{ra4} with the fit where the high-energy background is absorbed in the SOE. It is clear that neither the dispersion nor the linewidth is particularly sensitive to the parameters of the background, at least while these are close to minimizing $\langle \chi_{\rm R}^2 \rangle_Q$, and the SOE band centers we extract lie at the same energies as in the Ruddlesden-Popper compounds. By contrast, disallowing an additional high-energy background term drives the extracted SOE position up by approximately 10\% and the linewidth up by a value in excess of 30\%. Hence we conclude that the key difference in fitting this background is qualitative, i.e. in choosing to allow an additional high-energy contribution or not, and that including such a term offers a good, and reasonably unbiased, estimate of the intrinsic quantitative properties of the SOE. 

\begin{figure*}[t]
\centering
\includegraphics[width=0.88\textwidth]{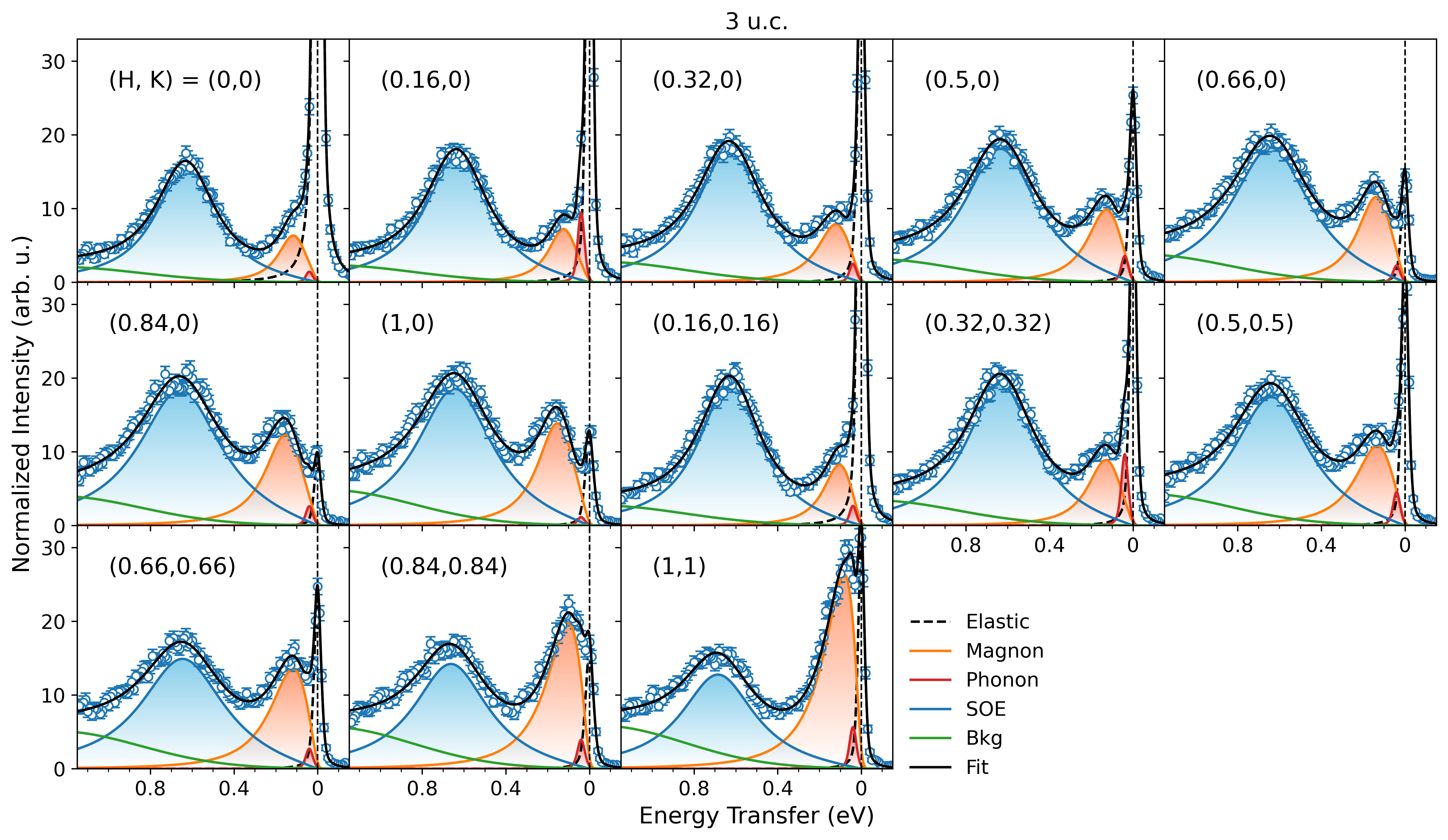}
\caption{\label{fig:SIO3fit} Fit of the Ir $L_3$-edge RIXS data at every momentum transfer for the 3 u.c.~sample. Circles show the RIXS data and the solid black line shows the best fit. The figure legend marks the individual components of the model fit.}
\end{figure*}

\begin{figure*}[t]
\centering
\includegraphics[width=0.88\textwidth]{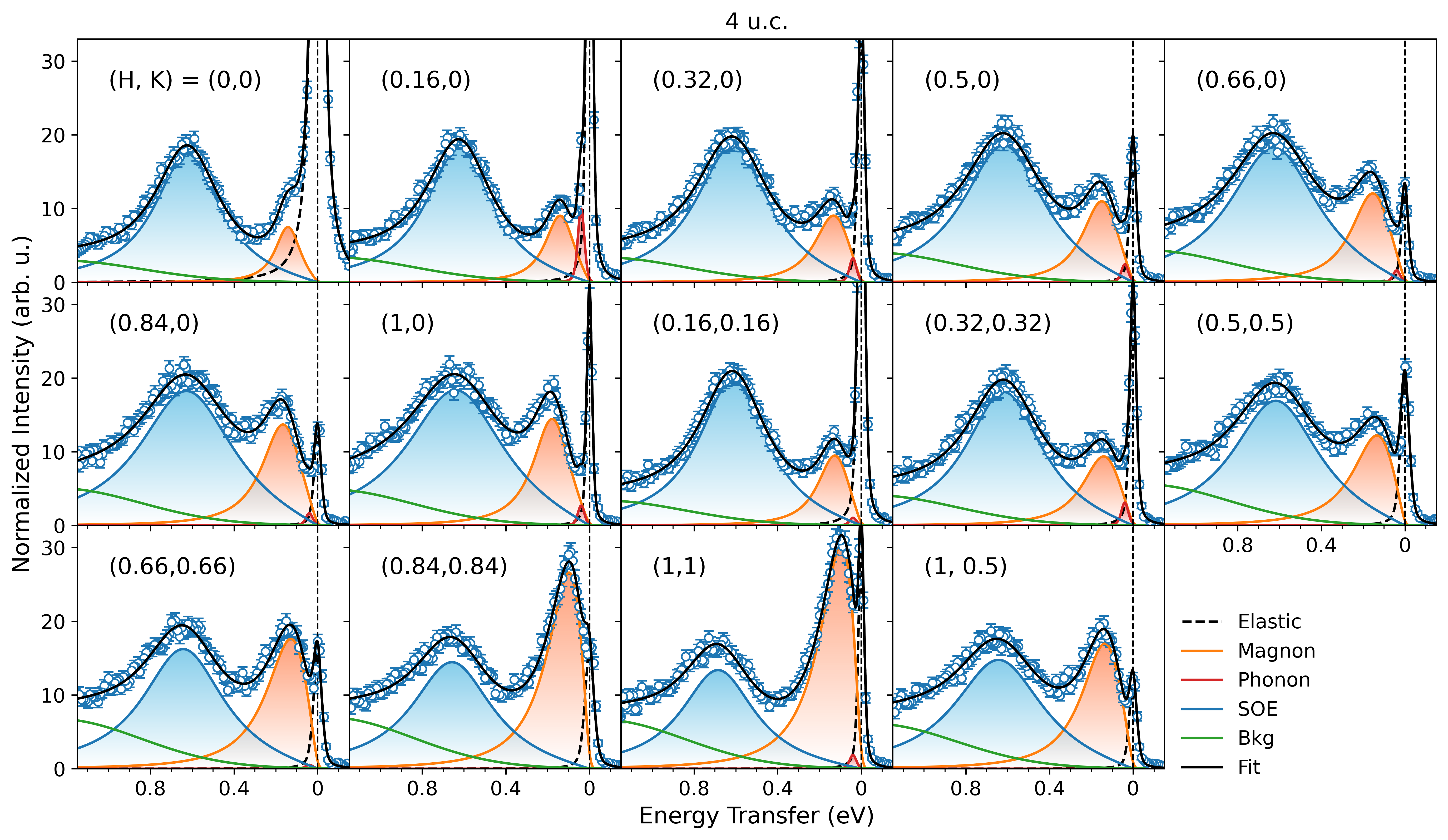}
\caption{\label{fig:SIO4fit} Fit of the Ir $L_3$-edge RIXS data for the 4 u.c.~sample, with notation as in Fig.~\ref{fig:SIO3fit}.}
\end{figure*}

\begin{figure*}[t]
\centering
\includegraphics[width=0.88\textwidth]{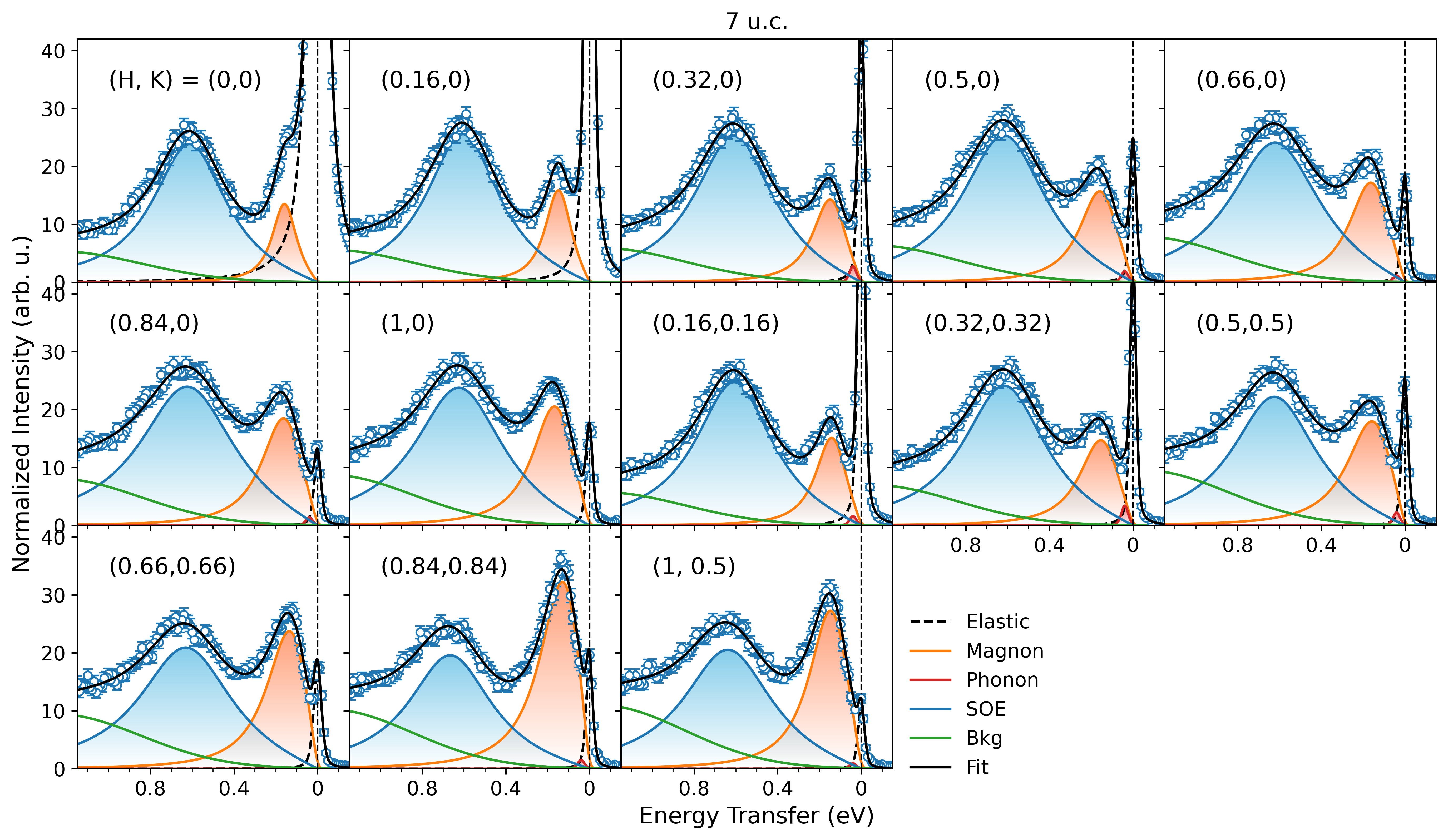}
\caption{\label{fig:SIO7fit} Fit of the Ir $L_3$-edge RIXS data for the 7 u.c.~sample, with notation as in Fig.~\ref{fig:SIO3fit}.}
\end{figure*}

\begin{figure*}[t]
\centering
\includegraphics[width=0.88\textwidth]{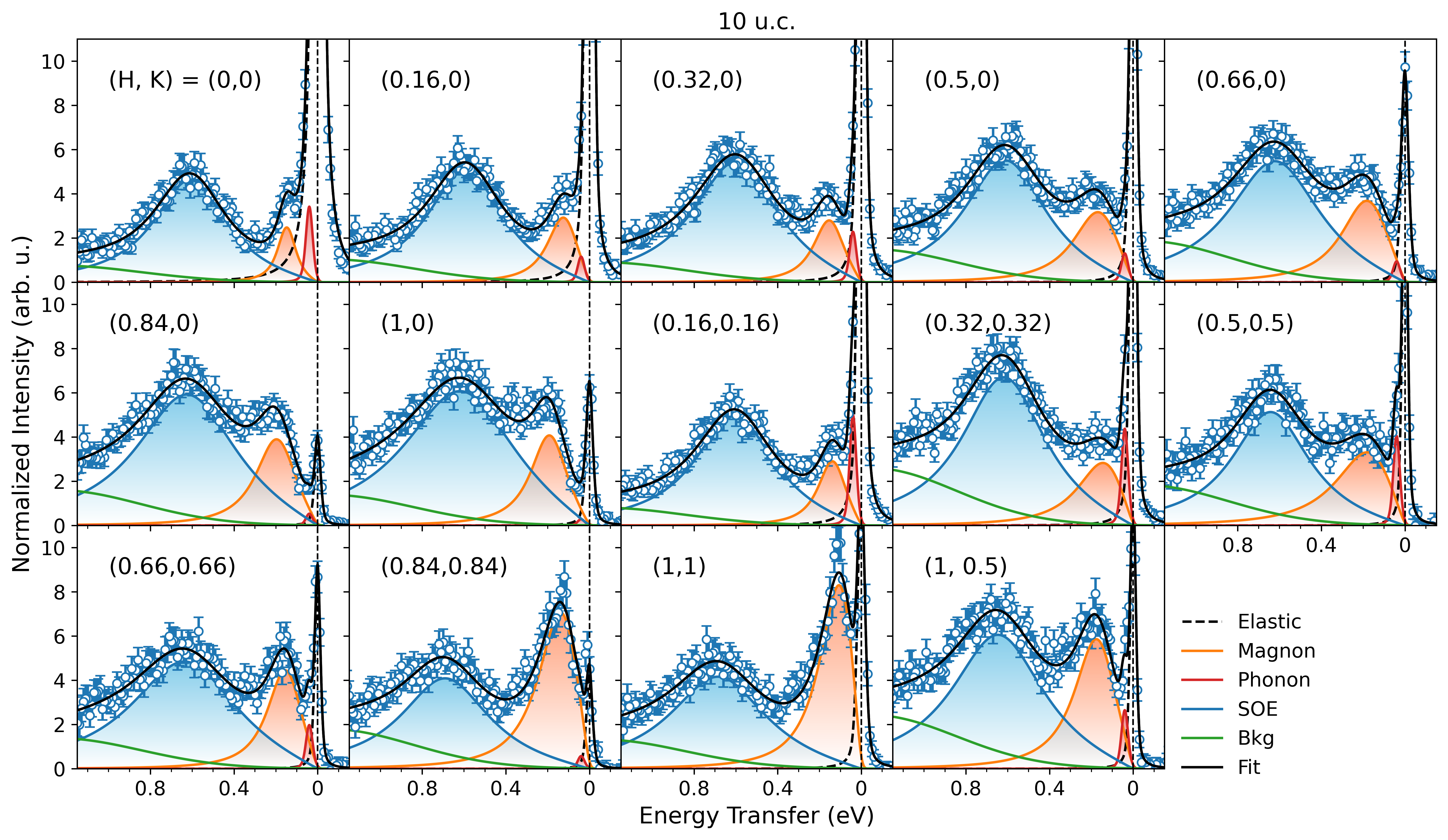}
\caption{\label{fig:SIO10fit} Fit of the Ir $L_3$-edge RIXS data for the 10 u.c.~sample, with notation as in Fig.~\ref{fig:SIO3fit}.}
\end{figure*}

Taking all of these contributions together [Eqs.~\eqref{eq:ff}], the total number of fitting parameters is $N_p = 10$, namely two for the elastic contribution, one for the phonon, three for the paramagnon, three for the SOE, and one for the high-energy background. In Figs.~\ref{fig:SIO3fit}-\ref{fig:SIO10fit}, we present the full Ir $L_3$-edge RIXS data for the 3, 4, 7, and 10 u.c.~samples for 13 or 14 momentum points along with the best fits of each spectrum to the five different components.

Based on this fitting procedure, the parameters of the DHO profiles for the paramagnon at each {\bf Q} are shown in Fig.~3 of the main text and for the SOE in Fig.~4. In the inset of Fig.~3(b) of the main text, we show how $\omega_{\rm i}$ obtained with this model differs from the actual peak position, $\omega_{\rm peak}$, corresponding to the maximum of the DHO profile. This quantity may be expressed as a function of $\omega_{\rm i}$ and $\gamma_{\rm i}$ as \cite{Robarts}
\begin{equation} \label{maxpos}
\omega_{\rm peak} = \frac{1}{6} \sqrt{12 \omega_{\rm i}^2 - 6\gamma_{\rm i}^2 + 6\sqrt{\gamma_{\rm i}^4 - 4\gamma_{\rm i}^2\omega_{\rm i}^2 + 16\omega_{\rm i}^4}}.
\end{equation}

\begin{figure}[t]
\centering
\includegraphics[width=0.94\linewidth]{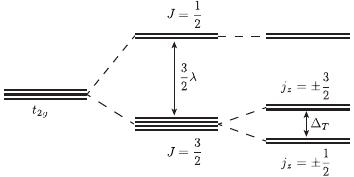}
\caption{\label{fig:tetragonal} Schematic representation of the energy-level structure of a $d$-orbital system in a cubic environment in the presence of a large spin-orbit coupling ($\lambda$). The $t_{2g}$ states split into a $J = 1/2$ doublet and a $J = 3/2$ quartet. In the presence of a tetragonal crystal field, $\Delta_{T}$, the quartet is further split according to the value of the $j_z$ quantum number. }
\end{figure}

\section{Comparison of heterostructure SOE{\lowercase {s}} with the AF insulator}
\label{ssoe}

\subsection{SOE in Sr$_2$IrO$_4$}

In Fig.~4(a) of the main text, we illustrate the very different dispersion of the SOE in our series of semimetallic samples when compared to the SOE in the AF insulator Sr$_2$IrO$_4$ \cite{Kim2012,Kim2014Excitonic}. Here we explain the extraction of the curve we show in Fig.~4(a) from the data of Ref.~\cite{Kim2012}. For transition-metal ions in a cubic environment, the six $t_{2g}$ orbitals lie below the four $e_g$ ones. A strong spin-orbit coupling splits the $t_{2g}$ manifold into $J = 1/2$ and 3/2 multiplets, as shown in Fig.~\ref{fig:tetragonal}, and the SOE is the excitation corresponding to $d$-$d$ transitions between these multiplets. A tetragonal distortion of the IrO$_6$ octahedra splits the $J = 3/2$ manifold into two Kramers doublets with energetic separation $\Delta_{T}$, which correspond to the $j_z = \pm 1/2$ and $j_z = \pm 3/2$ states (Fig.~\ref{fig:tetragonal}).
  
In bulk Sr$_2$IrO$_4$, $\Delta_{T} \simeq 140$ meV and the two branches are located around 550 meV and 700 meV. The 550 meV branch has stronger intensity at near-normal incidence and the 700 meV branch at near-grazing incidence \cite{Kim2014Excitonic}. Because our Ir $L_3$-edge RIXS data were obtained near grazing incidence, we compare our result with the high-energy branch of the SOE in Sr$_2$IrO$_4$, taking from Fig.~4(a) of Ref.~\cite{Kim2012} the dispersion of the mode we depict in Fig.~4(a) of the main text. 
In contrast to Sr$_2$IrO$_4$, where the SOE is sufficiently sharp that the two branches are resolved clearly, the significantly higher linewidth in \SIO~means that $\Delta_{T}$ is not resolved, and hence it contributes further to the appearance of a very broad SOE. 

\subsection{SOE in heterostructures}

The energetics of the SOE in our heterostructured samples can be affected by their more complex lattice structure, because the $d$-electron energy levels depend on both the geometry of the individual IrO$_6$ octahedra and on their relative rotations or bucklings \cite{Nie2015a,PhysRevLett.119.077201,Meyers2019}. In Sr$_2$IrO$_4$, where the octahedra are extended apically by $+9$\%, $\Delta_T = 140$ meV and Eqs.~(2-3) of Ref.~\cite{MorettiSala2014} yield a level splitting of 98 meV. In our heterostructures, the average out-of-plane lattice parameter is 3.98~\AA, compared to the in-plane value of 4.005~\AA, which corresponds to an apical compression of $- 0.6$\% if these changes are ascribed only to octahedral bond lengths, and not to rotation or buckling effects. By assuming an approximate linear scaling we obtain $|\Delta_T| < 10$ meV, whence the level splitting is less than 7 meV and cannot be resolved by RIXS. More detailed studies of other bulk systems, such as CaIrO$_3$, and of other heterostructures would be helpful in elucidating the role of the lattice in the energetics of the SOE.

Despite their crude nature, these considerations lead us to three separate observations. First, the degree of structural distortion and hence the SOE linewidth is not guaranteed to be constant across our heterostructure series, which may have a role in the $m$-dependence visible in Fig.~4(c) of the main text. Second, considering the role of the lattice in iridate materials in more general terms, it is certainly possible that one consequence of octahedral geometry is the broadening of the $J = 3/2$ bands that drives the system from the insulating to the metallic regime. Finally, because the effect of a tetragonal distortion appears to be so minor in our heterostructured samples, the physics underlying the very broad SOE peaks we observe must have a different origin, which points again towards the intrinsic metallic broadening that is also a property of bulk \SIO.

\end{document}